\documentstyle[prl,aps,epsf]{revtex}
\begin{document}

\draft

\title{Aharonov-Bohm Effect for Quasiparticles around
a Vortex Line in a D-wave Superconductor}

\author{ A. S. Mel'nikov}
\address{Institute for Physics of Microstructures,
        Russian Academy of Sciences\\
         603600, Nizhny Novgorod, GSP-105, Russia}

\date{\today}
\maketitle
\begin{abstract}
On the basis of the Bogolubov--de Gennes theory we develop an analytical
description of low-energy extended quasiparticle states around an
isolated flux line in a superconductor with gap nodes.  The wavefunctions
of these excitations and the corresponding density of states are shown to
be strongly influenced by the interaction with a pure gauge potential due
to the Aharonov--Bohm scenario.
\end{abstract}
\narrowtext
\twocolumn
Understanding the nature of low energy quasiparticle (QP) excitations
in isolated vortices and vortex lattices of type-II superconductors is of
considerable importance since for low temperatures $T$ these excitations
impact on various static and dynamic properties.
The flux lines are known to affect the low energy QP states
through the following mechanisms:
(i)~QP scattering on the gap inhomogenuity in vortex cores
(which is known to result in formation of localized core states in
s-wave compounds) \cite{degen};
(ii)~QP scattering on the potential
proportional to the superfluid velocity ${\bf V}_s$ \cite{cleary};
(iii)~long-range magnetic field effects which
are responsible for a finite curvature of a quasiclassical trajectory in
a vortex lattice \cite{gor};
(iv)~interaction with a pure gauge potential due to the Aharonov--Bohm
(AB) scenario \cite{cleary,fr1}.
For conventional s-wave superconductors these
mechanisms have been studied for several decades and now one may conclude
that the physical picture of the electronic structure of the mixed state
in s-wave systems is rather clear.
The situation is dramatically different in superconductors with gap
nodes where the vanishing pair potential in the nodal directions results
in qualitative changes in quantum mechanical motion of QPs.  The
interest to these fundamental issues is stimulated by recent experimental
observations of unconventional behavior of QP excitations in the mixed
state of high-$T_c$ compounds
\cite{moler,revaz,krishana,aubin,YBCO,hoogen,pan} where the
dominating order parameter is believed to be of d-wave symmetry.

In this Letter we focus on the quantum mechanical effects
caused by the interaction of QPs with pure gauge potentials induced by
flux lines. The main goal of our theoretical analysis is to show that
contrary to the s-wave case the AB effect plays a crucial
role in the behavior of low-lying QP states in d-wave compounds.
Let us consider a single isolated vortex line
which carries the flux quantum $\phi_0=\pi\hbar c/|e|$.
Due to the Meissner effect a magnetic field ${\bf H}$ is screened at the
London penetration depth $\lambda_L$ and at large distances from the
vortex center $r\gg\lambda_L$ the vector potential takes the form:  ${\bf
A}= \phi_0 [{\bf z}_0,{\bf r}]/(2\pi r^2)$, where $[{\bf z}_0,{\bf r}]$
is a vector product of vectors ${\bf z}_0$ and ${\bf r}$, and ${\bf z}_0$
is a unit vector chosen along the vortex axis.  Such a vector potential
can not be excluded from the Bogolubov -- de Gennes (BdG) equations
(describing the quantum mechanics of QPs) using any single-valued gauge
transformation. Note that within the standard Schroedinger theory for a
particle with a charge $e$ an AB solenoid with a magnetic flux
$\Phi=\phi_0$ is known to be the most effective scatterer \cite{ab}.
However one can see that for s-wave superconductors the effect of the
AB potentials on the QP density of states
(DOS) should be rather small.  The point
is that the size of the semiclassical wavepacket propagating outside the
flux tube (i.e., in the region $r\gg\lambda_L$)
appears to be much less than the impact
parameter. As a result, the interference effects are small and can be
neglected. As we see below, such a conclusion is no more valid if we
consider superconductors with gap nodes. Hereafter we assume
Fermi surface (FS) to be
two-dimensional (2D), which is appropriate to high-$T_c$ superconductors,
and take the gap function in the form $\Delta_d=2\Delta_0 k_xk_y/k_F^2$
(the $x$ axis is taken along the [110] crystal direction and thus makes
an angle $\pi/4$ with the $a$ axis of the $CuO_2$ planes).
The nature of low energy QP excitations
in the vortex state of d-wave systems
attracted recently a great deal of attention (see
\cite{gor,fr1,vo,so,wa,ma1,le,ma,fr,kopnin,ichi,me,an,ki,mr,ye}
and references therein).
In contrast to conventional superconductors, the DOS at low
energies $\varepsilon$ in d-wave systems is dominated by contributions
which come from the regions far from the vortex cores \cite{vo} and
associated with extended QP states with momenta close to the nodal
directions. This conclusion based on the semiclassical approach has been
confirmed by the recent numerical analysis \cite{fr} of the BdG
equations for a single isolated vortex line
(in the limit $\lambda_L\rightarrow \infty$).
Note that the calculations presented in \cite{fr} also point to the
absence of truly localized core states or any resonant levels in the
pure d-wave case though such states were observed in numerical
simulations \cite{ma}.

Let us start with some qualitative arguments which indicate a
significance of pure gauge potentials in the quantum mechanics
of extended QP excitations.
In the homogeneous state the low energy excitations
(confined to one of the gap nodes) are described by a Dirac-like spectrum:
$\varepsilon^\pm=
\pm\hbar\sqrt{V_F^2q_{\perp}^2+V_\Delta^2 q_{\parallel}^2}\ ,
$
where $V_F$ is the Fermi velocity, and
$(q_{\parallel},q_{\perp})$ defines a coordinate system whose
origin is at the node, with $q_{\perp}$ ($q_{\parallel}$) normal
(tangential) to the FS. The $V_\Delta$ value determines the
gap slope near the node and for our choice of the gap function
is given by the expression ${V_\Delta=2\Delta_0/(\hbar k_F)}$.
One can separate the following length scales
for QP wavefunctions: an atomic length scale $k_F^{-1}$ and two
characteristic wavelengths of a slowly varying envelope ${\it
l}_\perp\sim q_\perp^{-1}\sim \hbar V_F/\varepsilon$,  ${\it
l}_\parallel\sim q_\parallel^{-1}\sim \hbar V_\Delta/\varepsilon$.  The
length scales ${\it l}_\perp$ and ${\it l}_\parallel$ determine the size
of the semiclassical wave packet, which appears to diverge in the low
energy limit. Such a divergence is responsible for a crucial
role of quantum mechanical effects in QP motion and, in particular, for
the extremely large AB scattering amplitude.
This conclusion is in a good agreement with the recent calculations of
the QP band spectrum \cite{fr1,me,ki,mr} and thermal conductivity
\cite{ye} in the regime $H_{c1}\ll H\ll H_{c2}$.
Our further consideration of the low-energy QP states
($|\varepsilon|\ll \Delta_0$) with momenta locked to the nodes is based
on the essentially 2D quantum mechanical model proposed by Simon and Lee
\cite{le}.  Let us introduce the gap function describing the mixed state
of a d-wave superconductor in the form $\Delta({\bf k},{\bf
R})=\Delta_d({\bf k})\Psi({\bf R})$, where $\Psi({\bf R})=fexp(i\chi)$ is
the order parameter used in the Ginzburg-Landau theory.  In the limit
$k_F\xi\gg 1$ ($\xi$ is the coherence length) one can divide out the fast
oscillations on a scale $k_F^{-1}$ in the particle-like and hole-like
parts of the wavefunction $(u,v)=(\tilde u,\tilde v)exp(i{\bf k}_F{\bf
r})$ and simplify the nonlocal off-diagonal terms in BdG equations.
Omitting details of the derivation, we start from the BdG equations
linearized in gradient terms for quasiparticles with momenta close to a
certain gap node direction (e.g., ${\bf k}_1=(k_F,0)$):  ${\hat H\hat
g=\varepsilon\hat g}$, where $\hat g=(\tilde u exp(-i\chi),\tilde v)$,
and the Hamiltonian $\hat H$ outside the vortex cores takes the form:
\begin{equation}
\label{arb}
\hat H= \hat H_0+ |e|\varphi \ ,
\end{equation}
\begin{equation}
\label{arb1}
\hat H_0=-\hbar V_F \hat\sigma_z\left(i\frac{\partial}{\partial x}-
\frac{\chi^{\prime}_x}{2} \right)-
\hbar V_\Delta\hat\sigma_x\left(i\frac{\partial}{\partial y}-
\frac{\chi^{\prime}_y}{2} \right)
\end{equation}
Here $\hat\sigma_x,\hat\sigma_z$ are the Pauli matrices,
the scalar potential $\varphi=MV_FV_{sx}/|e|$ results in the Doppler
shift of the QP energy, $M$ is the electron effective mass, $V_{sx}$ and
$V_{sy}$ are the components of the superfluid velocity
${\bf V}_s=V_{sx}{\bf x}_0+V_{sy}{\bf y}_0$, ${\bf x}_0$, ${\bf y}_0$,
${\bf z}_0$ are the unit vectors of the coordinate system with ${\bf
z}_0$ chosen along the $c$-axis.  For an isotropic FS such a
description is correct only for $\varepsilon\ll \Delta_0/(k_F\xi)$,
but the range of validity may become larger if the FS
is somewhat flattened at the nodes.  The single-valued
transformation of wavefunctions ($\hat g=(\tilde u exp(-i\chi),\tilde
v)$) used above was introduced previously in \cite{me,an}.
Note that other possible versions of single-valued gauge transformations
were discussed in \cite{fr1,mr}.  The superfluid velocity ${\bf
V}_s({\bf r})$ can be written as a superposition of contributions from
individual vortices situated at points ${\bf r}_i$:
$$
{\bf V}_s({\bf r})=\frac{\hbar}{2M}\sum_{i}
\frac{[{\bf z}_0,{\bf r}-{\bf r}_i]}{\lambda_L|{\bf r}-{\bf r}_i|}
K_1\left(\frac{|{\bf r}-{\bf r}_i|}{\lambda_L}\right)
\ ,
$$
where $K_1$ is the Mcdonald function (modified Bessel function) of the
first order.  Note that in this expression we neglect the effects of
nonlocal electrodynamics \cite{amin1} which can not perturb the wave
function $\hat g$ essentially in the limit $\varepsilon<\hbar
V_\Delta/\xi$ when the minimum characteristic wavelength ${\it
l}_{min}\sim \hbar V_\Delta/\varepsilon$ exceeds the coherence length
$\xi$. One can see that our equations are analogous to the ones
describing the quantum mechanical motion of a massless Dirac particle
with a charge $|e|$ in
the `vector potential' ${\bf a}=-\phi_0\nabla\chi/(2\pi)$ of
AB solenoids and the scalar potential $\varphi$ of 2D
`electrical dipoles' screened at a length scale $\lambda_L$.  Both the
solenoids and the dipoles are positioned at ${\bf r}_i$. Each solenoid
carries the flux quantum $\phi_0$. The expression for a dipole moment
reads:  ${\bf P}=-0.5\hbar V_F {\bf y}_0/|e|$.  The Hamiltonian
(\ref{arb}) provides a simple tool for the study of 2D quantum mechanical
effects in the QP motion and has been recently used as a starting point
for both the analytical and numerical analysis of the band spectrum in
regular vortex arrays with a rather small intervortex distance $R_v\ll
\lambda_L$ \cite{fr1,me,mr}.

{\it Isotropic Dirac cone.}
Let us focus on the case of a single isolated vortex line. We start our
analysis from the most simple isotropic limit $V_F=V_\Delta=V$ (isotropic
Dirac cone) and show that the low energy QP states near each node are
strongly influenced by the pure gauge potential ${\bf a}$ due to the
AB scenario.  Indeed, if we neglect the potential of a 2D
`electric dipole' in (\ref{arb}), the scattering cross section
of a Dirac fermion in the AB potential appears to diverge for
$\varepsilon\rightarrow 0$:
${\displaystyle\frac{d\sigma}{d\theta}\propto \varepsilon^{-1}}$
\cite{alf,gerbert,hagen,gorni}.  Introducing a polar coordinate system
$(r,\theta)$ with the origin at the vortex center
and taking the order parameter phase in the form $\chi=\theta$
one can obtain the eigenfunctions of the Hamiltonian $\hat H_0$:
\begin{eqnarray}
\label{gm}
\hat g^{(1)}_m \propto (1+i \hat\sigma_x)\sqrt{\frac{k}{L}}
\left(e^{i m\theta}
J_{m+1/2}(kr)\atop sgn \varepsilon\  e^{i (m+1)\theta}
J_{m+3/2}(kr)\right)\ , \\
\label{fm}
\hat g^{(2)}_m\propto i\hat\sigma_ye^{-i\theta} \hat g^*_m \ ,
\end{eqnarray}
where $J_{\nu}$ is the $\nu$-th Bessel function,
$m$ is an integer,
$|\varepsilon|=\hbar V k$, and $L$ is the system size.
QP states confined near each node provide the following
contribution to the local DOS:
\begin{equation}
\label{dos}
N=\sum_{m}\frac{L}{2\pi}\int dk
 (|\tilde u_{m}|^2\delta (\varepsilon-\varepsilon_k)+
|\tilde v_{m}|^2\delta (\varepsilon+\varepsilon_k))\ ,
\end{equation}
where $\varepsilon_k=\pm\hbar V k$.
The wavefunctions $\hat g^{(1)}_m$
and $\hat g^{(2)}_m$ with $m\geq 0$ are regular at the origin.  The local
DOS corresponding to a set of these regular solutions vanishes in the
region $r<\hbar V/|\varepsilon|$ and approaches the value $N_\infty
\propto |\varepsilon|$ for $r\gg\hbar V/|\varepsilon|$ ($N_\infty$ is the
DOS in the absence of vortices).  The solutions with negative $m$ diverge
at the origin and are responsible for the formation of the nonzero DOS
near the vortex inside the domain $r<\hbar V/|\varepsilon|$. Thus, the
residual DOS $N_0$ (at $\varepsilon=0$) is also determined by the
contributions from the states with negative $m$. The crucial role in the
formation of the residual DOS is played by the states $g^{(1)}_{-1}$ and
$g^{(2)}_{-1}$ which appear to decay most slowly from the vortex center
(as $r^{-1/2}$) in the limit $\varepsilon\rightarrow 0$. Using
(\ref{dos}) one can obtain the following contribution to the residual
local DOS:  $N_0\sim (\hbar V r)^{-1}$. Let us emphasize
that this contribution appears to be nonzero even in the large-{\bf r}
domain ($r>\lambda_L$) in sharp contrast to the behavior expected on the
basis of the semiclassical model which takes account of the Doppler term
$|e|\varphi$ and neglects the AB effect.

We now proceed with the analysis of the effect of the Doppler term
in the Hamiltonian (\ref{arb}) on the behavior of wavefunctions in
the small-{\bf r} domain ($r<\lambda_L$).
For low energies $\varepsilon\ll \hbar V/\lambda_L$ the regular solutions
(\ref{gm},\ref{fm})
with $m\geq 0$ are only weakly influenced by the potential of the
screened `electric dipole'.  On the contrary, for the wave functions
(\ref{gm},\ref{fm}) with negative $m$ (which diverge at $r\rightarrow 0$)
the scalar potential $\varphi$ can not be considered as a small
perturbation and essentially modifies the solution in the domain
$r<\lambda_L$. In the low energy limit $\varepsilon\ll \hbar V/\lambda_L$
the solutions in this ${\bf r}$-domain can be written in the form:
\begin{eqnarray}
\label{st1}
\hat g^{(1)}_\mu=(1+i \hat\sigma_x) r^{\mu+1/2}
exp(-i\theta(1+\hat\sigma_z)/2)\hat G_\mu (\theta) \ ,
\\
\label{st2}
\hat g^{(2)}_{\mu}\propto i\hat\sigma_ye^{-i\theta} \hat g^*_{\mu} \ ,
\end{eqnarray}
where the equation for $\hat G_\mu (\theta)$ reads:
\begin{equation}
\label{theta}
i\frac{\partial}{\partial \theta}\hat G_\mu+
(1+\mu)\hat\sigma_z \hat G_\mu+\frac{sin\theta}{2}\hat\sigma_x
\hat G_\mu =0\ .
\end{equation}
A set of discrete quantum numbers $\mu$ is determined by the condition:
$\hat G_\mu (\theta)=\hat G_\mu (\theta+2\pi)$.
In the large-$|\mu|$ limit the $\mu$ values can be calculated using
the quasiclassical quantization rule:
\begin{equation}
\int\limits_0^{2\pi}\sqrt{(1+\mu)^2+\frac{1}{4}sin^2\theta}d\theta=
2\pi n\ ,
\end{equation}
where $n$ is an integer. The residual DOS near the vortex center is
dominated by the states (\ref{st1},\ref{st2}) with $\mu<-1/2$, which are
characterized by the power divergence at $r\rightarrow 0$.  The
divergence should be cut off due to the matching with the solution inside
the core, which results in a strong mixing of QP wavefunctions
characterized by different $\mu$ values and corresponding to all four
nodes. Outside the core the most slowly decaying wavefunctions correspond
to the value $\mu=-1$ and can be found exactly using Eq.(\ref{theta}):
\begin{eqnarray}
\label{gms}
\hat g^{(1)}_{-1} = C (1+i \hat\sigma_x) r^{-1/2}
\left(e^{-i\theta} cos\gamma\atop -i sin\gamma\right) \ ,
\\
\label{fms}
\hat g^{(2)}_{-1}= i\hat\sigma_ye^{-i\theta} \hat g^*_{-1} \ ,
\end{eqnarray}
where $\gamma=cos\theta /2$,
and the constant $C\propto L^{-1/2}$ is determined from the matching
with the solution in the large-${\bf r}$ domain ($r>\lambda_L$).
One can see that for intermediate distances $\xi\ll r\ll \lambda_L$
these states provide a dominating contribution to the residual local
DOS $N_0\sim (\hbar V r)^{-1}$, which coincides with the one
predicted on the basis of the semiclassical model \cite{vo}.
Obviously the latter conclusion is no more valid
near the vortex core, where the admixture of the solutions
decaying more rapidly from the vortex axis becomes substantial
and can result in a narrow DOS peak similar to the one observed in
high-$T_c$ compounds \cite{YBCO,hoogen,pan}.

Despite of a possible formation of such a peak structure
the resulting low-energy states are not truly localized.
The wavefunctions $\hat g^{(1)}_{-1}$ and $\hat g^{(2)}_{-1}$
can be considered as leading terms in a large distance
($r\gg\xi$) asymptotic expansion for the low-lying states and, thus, are
responsible for the escape of QPs from the core.  These conclusions can be
supported by the analysis of the generalized inverse participation ratio,
defined as \cite{fr}:
$$
\beta\propto L^2\frac{\int
(|\tilde u|^4+|\tilde v|^4) d{\bf r}}
{\left(\int (|\tilde u|^2+|\tilde v|^2) d{\bf r}\right)^2}\ .
$$
For slowly decaying solutions
$\hat g^{(1)}_{-1}$ and $\hat g^{(2)}_{-1}$  the $\beta$ value
appears to grow logarithmically as a function of an increasing system
size $L$. Such logarithmic behavior is in a sharp contrast to the
$\beta\propto L^2$ divergence expected for localized states.
The absence of the $\beta\propto L^2$ scaling is in a good agreement
with the numerical analysis \cite{fr} carried out in the limit
$L\ll \lambda_L$.

{\it Anisotropic Dirac cone.}
The physical picture suggested above can be generalized for the case of
an anisotropic spectrum $V_F\neq V_{\Delta}$. Such a generalization is of
particular importance for the understanding of the electronic structure
of vortices in high-$T_c$ cuprates, where an estimate based on
the results of thermal conductivity measurements \cite{thermo}
gives us a rather large anisotropy ratio: $\alpha=V_F/V_{\Delta}\sim 14$.
Taking an appropriate gauge transformation
$\hat g=e^{iS}\hat f$ one can choose the `vector potential' ${\bf a}$
in the Hamiltonian $\hat H_0$ in the form:
$$
{\bf a}=-\frac{\phi_0\alpha}{2\pi}
\frac{(-y{\bf x}_0+x{\bf y}_0)}{x^2+(\alpha y)^2} \ .
$$
Using the scale transformation ($\alpha y=\tilde y$, $x=\tilde x$)
the Hamiltonian $\hat H_0$ can be reduced to the isotropic one with
$V=V_F$. As a result, in the new coordinates the solution $\hat f$ in the
large-${\bf r}$ domain (where the Doppler shift is negligible)
can be written in the form (\ref{gm},\ref{fm}) with
$k=|\varepsilon|/(\hbar V_F)$. The small distance asymptotic
expansion can be
analysed analogously to the isotropic case. The most slowly decaying
solutions can be obtained from (\ref{gms},\ref{fms})
if we replace $\theta$ and $r$ by
$\tilde\theta=tan^{-1}(\tilde y/\tilde x)$ and
$\tilde r=\sqrt{\tilde x^2+\tilde y^2}$, respectively, and take
$$
\gamma=\frac{\alpha}{2\sqrt{\alpha^2-1}}
tan^{-1}(\sqrt{\alpha^2-1}cos\tilde\theta) \ .
$$
Outside the core the expression for the residual local DOS
(taking account of the contributions from all four nodes) reads:
\begin{equation}
\label{totdos}
N_0\sim \frac{1}{\hbar\sqrt{V_\Delta^2
x^2+V_F^2 y^2}}+ \frac{1}{\hbar\sqrt{V_\Delta^2 y^2+V_F^2 x^2}} \ .
\end{equation}
The anisotropy of the Dirac cone results in the angular dependence of the
local DOS which exhibits a fourfold symmetry with the maxima along the
nodal directions (see Fig.1).

{\it Multi-quanta flux structures.}
The generalization of the above analysis to the case
of a multi-quanta vortex with a winding number $M_\phi$
is straightforward.
For a vortex carrying an odd number $M_\phi$ of the flux quanta
the effect of the AB potential on the local DOS
outside the flux tube ($r>\lambda_L$) is the same as for a singly
quantized vortex.  On the contrary,
a vortex with an even number $M_\phi$ does
not cause the AB interference and, thus, can not provide the
slowly decaying contribution ($N_0\propto r^{-1}$)
to the residual DOS discussed above.
The most direct way to observe this odd-even effect is to consider a
hollow cylinder with a trapped magnetic flux $\Phi$.
The AB mechanism is the cause of an oscillating
dependence $N_0(\Phi)$, which should be observable by any
experimental technique sensitive to the residual DOS.
The large odd-even effect in the residual DOS is specific for
superconductors with gap nodes and can be considered
as a test for d-wave pairing.
Obviously, the magnitude of such odd-even effect is strongly influenced by
finite lifetime and temperature effects. One can expect that
these mechanisms should suppress the slowly decaying contribution
(\ref{totdos}) to the local DOS in the domain
$r>\hbar V_F/max[\Gamma,T]$, where $\Gamma$ is a scattering rate.

To sum up, we have discussed an extremely important role of pure gauge
potentials in the behavior of low energy QP excitations in the
mixed state of superconductors with gap nodes.  Within the BdG theory
linearized in gradient terms we have considered the peculiarities of
extended QP states for isolated flux lines.
Such a model allowed to develop a 2D quantum mechanical description of
these states, taking account of both the Doppler shift of the QP energy
and the Aharonov--Bohm effect. It is hoped that the physical picture
considered in this Letter can provide a starting point for the analysis
of static and dynamic properties of the mixed state in various d-wave
systems, including, probably, high-$T_c$ copper oxides.

I am indebted to  I.D.Tokman, D.A.Ryndyk, and A.A.Andronov
for useful discussions and to J.Ye for correspondence.
This work was supported, in part, by the
Russian Foundation for Fundamental Research (Grant No. 99-02-16188).

\begin{figure}[h]
\leavevmode
\epsfysize=6cm
\epsfbox{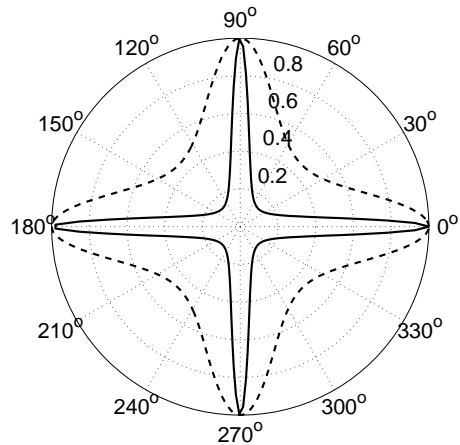}
\caption{
The angular dependence of the residual DOS
$N_0(\theta)/N_0(\theta=0)$ in polar coordinates for
$\alpha=5$ (dashed line) and $\alpha=20$ (solid line).}
\label{fig1}
\end{figure}
\end{document}